\documentclass[10pt]{article}

\usepackage{amsmath}
\usepackage{amssymb}

\usepackage{graphicx}

\usepackage{cite}

\usepackage{color} 

\usepackage{setspace} 
\usepackage[labelfont=bf,labelsep=period,justification=raggedright]{caption}

\makeatletter
\renewcommand{\@biblabel}[1]{\quad#1.}
\makeatother

\date{}

\begin{document}

\begin{flushleft}
{\Large
\textbf{Multiband Semimetallic Electronic Structure of Superconducting Ta$_2$PdSe$_5$}
}
\\
David J. Singh$^{1,\ast}$
\\
\bf{1} Materials Science and Technology Division,
Oak Ridge National Laboratory, Oak Ridge, Tennessee, USA
\\
$\ast$ E-mail: singhdj@ornl.gov
\end{flushleft}

\section*{Abstract}

We report the electronic structure and related properties of the superconductor
Ta$_2$PdSe$_5$ as determined from density functional calculations.
The Fermi surface has two disconnected sheets,
both derived from bands of primarily
chalcogenide $p$ states. These are a corrugated hole cylinder and
and a heavier complex shaped electron sheet.
The sheets contain 0.048 holes and a compensating number of electrons
per formula unit, making the material a semimetallic superconductor.
The results support the
presence of two band superconductivity, although a discrepancy in the
specific heat is noted. This discrepancy
is discussed as a possible consequence of Pd deficiency in samples.


\section*{Introduction}

Superconductivity with very large upper critical fields
has been reported in several layered compounds
$M_2$Pd$_x$$Ch_5$, $M$=Nb,Ta and $Ch$=S,Se.
\cite{lu-ta2pds5,zhang-nb2pds5,jha-nb2pds5,zhang-ta2pdse5,zhou-nb2pds5,
khim-nb2pdse5,yu-nb2pds5,niu-nb2pds5}
The upper critical field
is reported to  exceed an estimate of the Pauli limit \cite{clogston,werthamer}
for
Ta$_2$Pd$_{0.92}$S$_5$. \cite{lu-ta2pds5}
This has been discussed both in terms of spin-orbit effects,
\cite{lu-ta2pds5,zhou-nb2pds5,khim-nb2pdse5}
and multiband superconductivity.
\cite{singh-ta2pds5,goyal,zhang-ta2pdse5}

An important chemical feature of these compounds is that one of the
Pd sites in the unit cell is chemically less favorable than
the other, leading to a tendency for Pd deficiency.
\cite{lu-ta2pds5,singh-ta2pds5}
It was argued that, although the
average crystal structures are centrosymmetric,
this Pd deficiency can lead to local inversion symmetry breaking,
which with heavy elements such as Ta or Pd in the structure can lead to
enhancement of the upper critical fields from spin-orbit scattering.
\cite{lu-ta2pds5,zhou-nb2pds5,khim-nb2pdse5}
The fact that the samples invariably have large number of Pd vacancies
is consistent with this view.
In this regard, the metal atoms occur on one dimensional chains in the
crystal structure, which has been emphasized in relation to spin-orbit
effects. \cite{khim-nb2pdse5}

The large upper critical fields have also
been discussed in terms of multiband superconductivity.
In particular, it was noted that the electronic structure of Ta$_2$PdS$_5$
has two main bands, and in addition there is evidence
for strong coupling, which together could also account for the high
upper critical fields.
\cite{singh-ta2pds5}
Multiband superconductivity has also been discussed in the related
compounds Nb$_3$Pd$_{0.7}$Se$_7$,
\cite{zhang-nb3pdse7}
and Ta$_4$Pd$_3$Te$_{16}$. \cite{singh-ta4pd3te16}
The band structures do show multiple Fermi surfaces consistent with this
view and a recent report of the specific heat of Nb$_2$Pd$_x$S$_5$
with different applied fields is consistent with two different superconducting
gaps. \cite{goyal}

One way forward in understanding the high upper critical fields and
other properties of these materials is through the systematic property
variation between different compounds
and samples with different stoichiometry
in relation to band structure calculations.
Recently, Zhang and co-workers, \cite{zhang-ta2pdse5}
have reported a new system, Ta$_2$PdSe$_5$, which is superconducting
with critical temperature $T_c$=2.5 K, and has a low temperature
dependence of the specific heat consistent with multiband
superconductivity.
This newly reported compound has an upper critical field
$H_{c2}(0)$=15.5 T, for a ratio $H_{c2}(0)/T_c$= 6.2, similar
to the other compounds in the family.
The purpose of the present paper is to 
elucidate the electronic structure of this compound in relation
to experiment.

\section*{Results}

\begin{figure}[!ht]
 \includegraphics[width=\columnwidth]{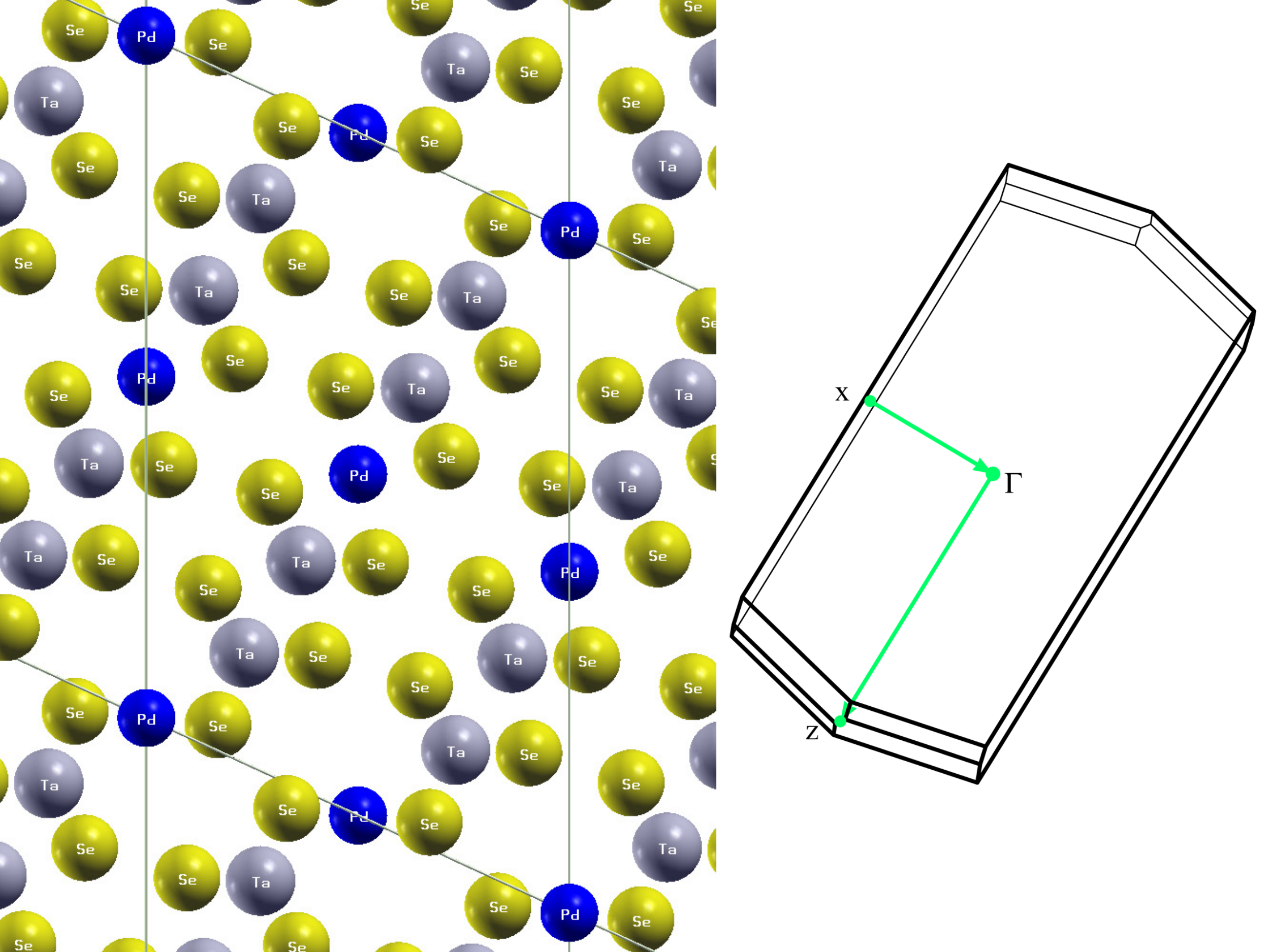}
\caption{{\bf Crystal structure of Ta$_2$PdSe$_5$ with the relaxed atomic
positions (left) and Brillouin zone (right).}
The crystal structure is viewed along the short $c$-axis, which is the
direction of the metal chains.  The conventional
unit cell is indicated. The colored lines in the
zone indicate the {\bf k}-point path in the band structure plot.}
\label{fig:struct}
\end{figure}

The calculated atomic positions from total energy minimization at the
experimental lattice parameters are given
along with the bond valence sums in Table \ref{tab:struct}.
The resulting structure is depicted in Fig. \ref{fig:struct}.
It follows the structure type as described by Squattrito and 
co-workers. \cite{squattrito}
It is qualitatively similar to that of Ta$_2$PdS$_5$,
described previously, \cite{singh-ta2pds5} and
in particular has a 4-fold coordinated Pd site connecting ribbons
of composition Ta$_4$PdSe$_{10}$.
This site is prone to vacancies in the sulfide. 
Deviation from the nominal stoichiometry has not been reported
in Ta$_2$PdSe$_5$, but the possibility of Pd vacancies associated with
the Pd2 site should be kept in mind.
The shortest metal-metal distances are along the $c$-axis,
forming the one dimensional metal chains that have been discussed
previously. \cite{khim-nb2pdse5} The metal atoms in these chains are
separated by $c$=3.117 \AA.
However, even from the point of view of metal-metal bonding, the
structure may not be best viewed as one-dimensional.
Specifically, there are comparable in-plane metal-metal distances.
For example, the Ta2-Pd2 distance is 3.229 \AA.
Importantly, the Pd2 site joins the ribbons, mentioned above,
and depending on the electronic hopping, the ribbons might or might
not be independent one dimensional objects from an electronic point of view.
The bond valence sums deviate substantially from nominal ionic values, except
for Ta.
The average Se value is 2.95. These deviations are indicative of
covalent bonding.

\begin{figure}[!ht]
  \includegraphics[width=\columnwidth]{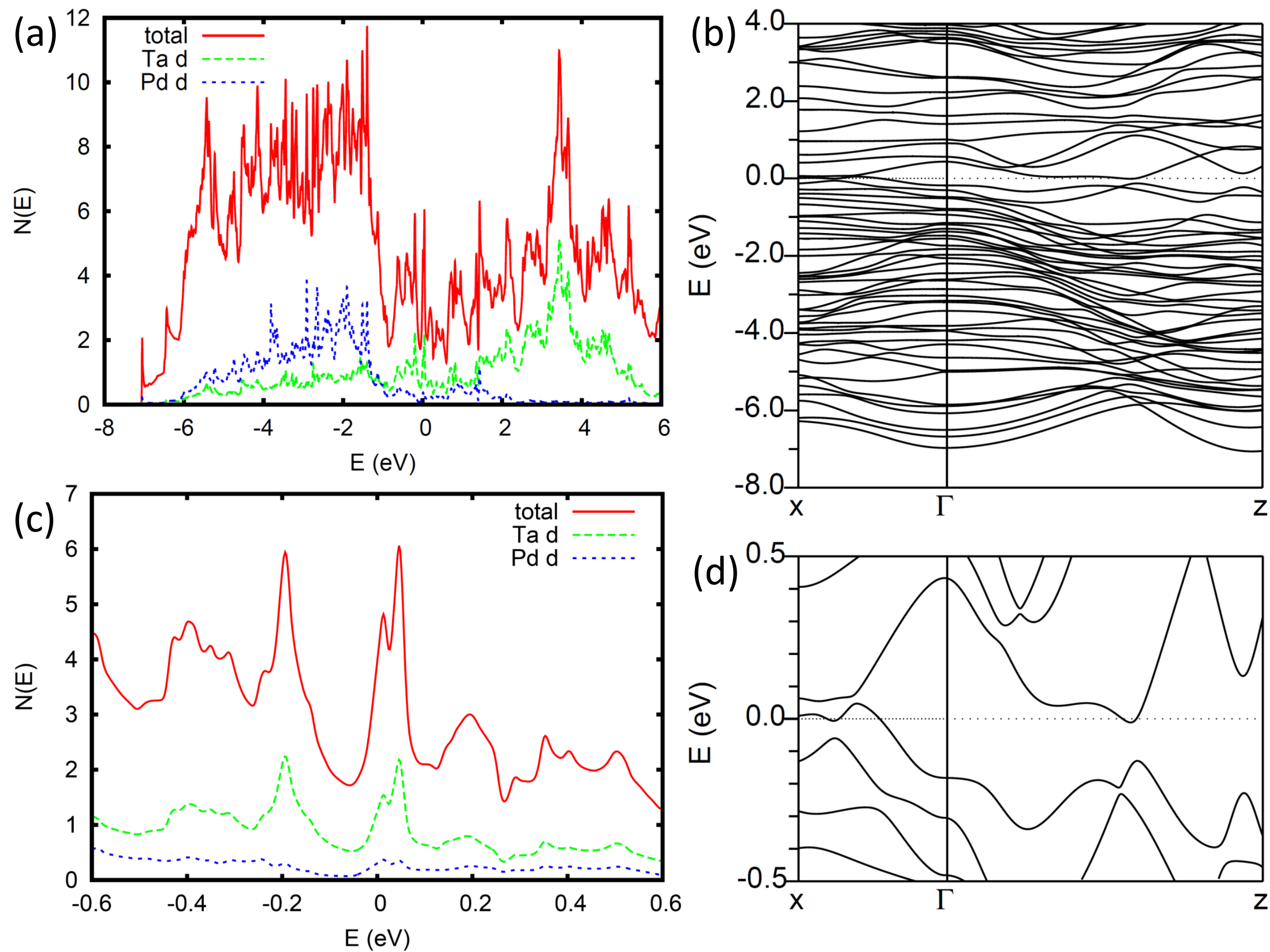}
\caption{{\bf Electronic structure of Ta$_2$PdSe$_5$.}
(a) The density of states on a per formula unit basis.
(b) The band structure along the directions given in Fig. \ref{fig:struct}.
(c) A blow-up of the density of states around $E_F$, which is at 0 eV
in all plots.
(d) The band structure near $E_F$.
}
\label{fig:dos}
\end{figure}

The calculated electronic density of states and projections
of Ta and Pd $d$ character are shown in Fig. \ref{fig:dos}, along
with the band structure.
As may be seen, the Ta $d$ states occur mainly above the
Fermi level, $E_F$, while the Se $p$ states and
Pd $d$ states are mainly below $E_F$.
The nearly fully
occupied Pd $d$ states
argue against the presence of spin-fluctuations
associated with Pd.
In any case, a high degree of covalency is evident both from the
Ta (Se) contributions below (above) $E_F$ and from the width of
the corresponding regions of the density of states.

The calculated density of states at the Fermi level is
$N(E_F)$=3.94 eV$^{-1}$ for both
spins on a per formula unit basis, and as
seen is mostly derived from Se $p$ states. The metal $d$ contributions are
0.86 eV$^{-1}$, 0.38 eV$^{-1}$, 0.23 eV$^{-1}$ and 0.37 eV$^{-1}$, for
Ta1, Ta2, Pd1 and Pd2, respectively, on a per metal atom basis,
for a total Ta $d$ contribution of 
1.24 eV$^{-1}$ and a total Pd $d$ contribution of 0.30 eV$^{-1}$
per formula unit. These low numbers place the materials very far away from
any transition metal magnetism. Furthermore, since superconductivity is
an instability of the Fermi surface of a metal, it is clear that the
electrons involved in pairing are in bands that are primarily derived
from chalcogen $p$ states, hybridized with mainly Ta1 $d$ states.

The bare electronic specific heat coefficient
from the calculated density of states
is $\gamma_{bare}$=9.6 mJ/(mol K$^2$).
This is very close to the measured $\gamma$=10.3 mJ/(mol K$^{2}$),
which is not expected, since there should be an enhancement
$\gamma$=(1+$\lambda$)$\gamma_{bare}$. 
Here $\lambda$ is roughly
the superconducting electron-phonon
coupling, $\lambda_{ph}$ in the absence of spin-fluctuations, and even
larger if there are substantial spin-fluctuation effects.
The low value inferred from comparison of the calculated $\gamma_{bare}$,
with the experimental $\gamma$ is not compatible with superconductivity.
In the experimental data of Zhang and co-workers
\cite{zhang-ta2pdse5}
the specific heat jump is found to be,
$\Delta C_e/\gamma T_c \sim$0.83, which is much smaller than the weak
coupling value of 1.43. Zhang and
co-workers note that this
could be due to a non-superconducting fraction in the sample.
A multiphase sample could also affect the inferred value of $\gamma$,
for example if one of the phases has a small $\gamma$.
Another likely explanation is that off-stoichiometry, e.g. Pd
vacancies is important. $E_F$ lies on the
leading edge near the top of a peak in the
density of states, which means that broadening due to disorder would
be expected to lower $N(E_F)$. Metal deficiency, if it lowers $E_F$
would also reduce $N(E_F)$.

\begin{figure}[!ht]
 \includegraphics[width=\columnwidth]{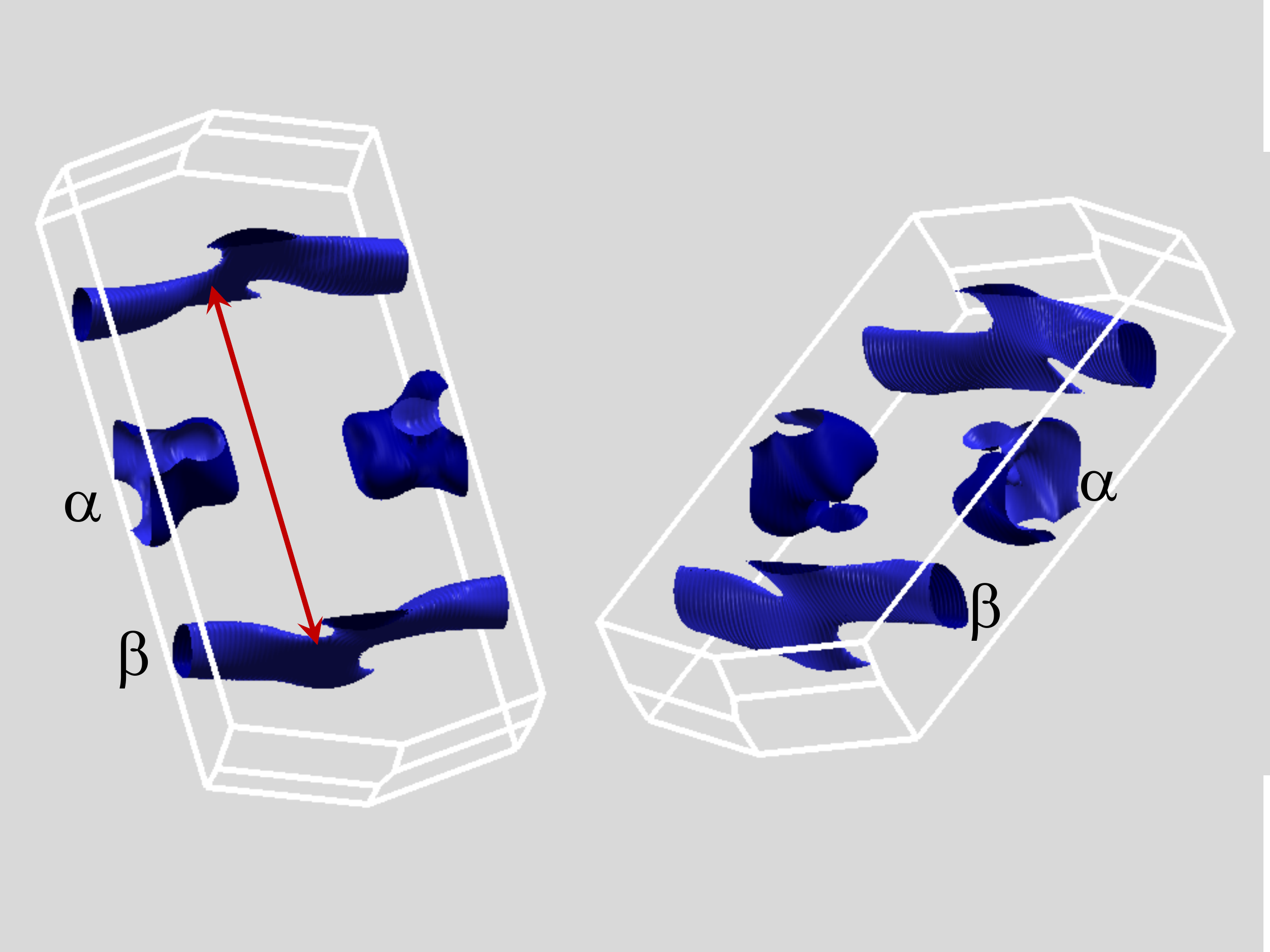}
\caption{{\bf Calculated Fermi surface of stoichiometric Ta$_2$PdSe$_5$.}
Two views are shown. The zone is as depicted in Fig. \ref{fig:struct}.
The two sheets, $\alpha$ and $\beta$ are as indicated. The red line
denotes the nesting vector for the $\beta$ sheet.
}
\label{fig:fermi}
\end{figure}

The Fermi surface is depicted in Fig. \ref{fig:fermi}.
It has two disconnected sheets.
Such a structure can be very favorable for superconductivity
with sign changes of the order between the sheets if there are
repulsive interactions, e.g. spin-fluctuations, that couple the sheets.
\cite{kuroki,mazin-spm,kuroki2,singh-na}
However, this is highly unlikely here because, as mentioned,
the bands near $E_F$ have relatively little transition metal character,
placing the material far from magnetic instabilities, while the
electron-phonon interaction is attractive.
The Fermi surface consisting of two disconnected sheets also allows
for multiband superconductivity, which was proposed as an explanation
for the high $H_{c2})(0)$ values
in this family of materials,
\cite{singh-ta2pds5}
following the general arguments of Gurevich and co-workers.
\cite{gurevich,gurevich2,gurevich3,hunte}
The two sheets of Fermi surface are a  hole sheet, $\alpha$, in the
form of a very corrugated cylinder around the
point labeled $x$ (Fig. \ref{fig:struct}) and a compensating
complex shaped electron surface, $\beta$,
consisting of interconnected
tube shaped sections.
These $\beta$ sheets are flat along the direction $\Gamma$-$z$, and
are therefore nested as shown by the line in Fig. \ref{fig:fermi}.
The $\alpha$ and $\beta$ Fermi surfaces each contain $\sim$0.048 of
the zone. Therefore there are 0.048 holes per formula unit 
(two formula units per cell) in the $\alpha$ pocket and a corresponding
number of electrons in the $\beta$ pocket.
Ta$_2$PdSe$_5$ is thus a semimetallic superconductor, as are
the Fe-based superconductors. \cite{singh-du}
The electron and hole Fermi surfaces
contribute $\sim$40\% and 60\%, respectively,
to $N(E_F)$, i.e. the electrons are heavier than the holes in
this compound.

Neither of the two sheets of Fermi surface is one dimensional
(a one dimensional surface associated with the metal chains would be
a flat section perpendicular to the $\Gamma$-z direction as labeled
in Fig. \ref{fig:struct}).
Both of the Fermi surfaces are associated with bands near an anticrossing.
This leads to complex shaped bands very near $E_F$, which may lead to
a particularly strong sensitivity to disorder. This may be present due to
Pd vacancies. The cylindrical hole sheet is associated with the inverted
band at the corresponding face of the zone. The band shape is interesting
from a topological point of view because of this band inversion and because
the position on the zone face yields a one pocket per zone, i.e. an odd
number of pockets.
In fact, the complex shapes of the Fermi surfaces in spite of the low
band fillings, are a direct consequence of the origin in band inversion
with spin-orbit
as discussed recently in the context of thermoelectric materials.
\cite{shi}

The conductivity tensor scaled by the scattering rate was
determined from the Fermiology within the constant
scattering time approximation. This assumes that the
scattering rates on the two sheets of Fermi surface are the same.
The resulting eigenvalues of $\sigma/\tau$ were
4.4x10$^{18}$ 1/($\Omega$ms) and
8.7x10$^{18}$ 1/($\Omega$ms) in the $a$,$b$, plane, and
3.10x10$^{19}$ 1/($\Omega$ms) along the $c$-axis direction.
Thus the conductivity is highest along the $c$-axis, which is
the direction of the ribbons and metal atom chains.
However, the anisotropy of $\sim$7 between the highest and lowest 
conductivity directions is not large enough to reasonably call this
a one dimensional conductor.

\section*{Discussion}

We report calculations of the structure and electronic properties
of the superconductor Ta$_2$PdSe$_5$. Perhaps not surprisingly,
the calculations show
this compound is in many respects very similar to
previously reported members of this family such as Ta$_2$PdS$_5$.
The calculated density of states is only slightly smaller than
the value inferred from the experimental specific heat, which
is not expected for a superconductor. This discrepancy, which
might be due to Pd vacancies in the samples, suggests further study.

Importantly, Ta$_2$PdSe$_5$ is an anisotropic three dimensional
metal with two disconnected compensating sheets of Fermi surface. This is
consistent with the explanation that the high upper critical fields
in these compounds are a manifestation of multiband superconductivity.
The calculations do not support the view that this material is
reasonably described as a one dimensional metal.
Both the Fermi surface sheets occur from bands near anticrossings.
This leads to a topological structure for the band forming the
hole cylinder.

Multiband superconductivity is a much discussed subject,
\cite{gurevich,suhl,kondo,zehetmayer}
but has a limited number of established examples. These include
unconventional superconductors, particularly the Fe-based
superconductors, \cite{hunte}
and the electron-phonon superconductor MgB$_2$.
\cite{koshelev,mazin-2gap}
Ta$_2$PdSe$_5$ and the related chalcogenides provide a new
family of materials exhibiting multiband electron-phonon superconductivity.
The relatively accessible values of the upper critical fields and
the variety of compounds in this family should make these compounds
particularly useful for studying multiband superconductivity.

\section*{Methods}

The electronic structure calculations were performed within density
functional theory using the generalized gradient approximation of
Perdew, Burke and Ernzerhof (PBE). \cite{pbe}
The calculations were performed using the general potential
linearized augmented planewave (LAPW) method,
\cite{singh-book} as implemented in the
WIEN2k code. \cite{wien2k}
The standard LAPW linearization was employed with local orbitals
\cite{singh-lo}
to treat the semicore states, specifically the 5$s$ and 5$p$ states of
Ta, the 4$p$ states of Pd and the 3$d$ states of Se.
LAPW sphere radii of 2.4 bohr, 2.3 bohr and 2.3 bohr were used
for Ta, Pd and Se, respectively.

Well converged basis sets were employed. The planewave
cutoff, $k_{max}$ was determined by $R_{min}k_{max}$=9,
where $R_{min}$=2.3 bohr was the smallest sphere radius.
The crystal structure was based on the experimental monoclinic
lattice parameters, in space group 12. \cite{zhang-ta2pdse5} The standard
$C2/m$ setting was used for convenience, similar to prior
calculations for the sulfide, Ta$_2$PdS$_5$.
\cite{singh-ta2pds5}
The internal atomic coordinates were determined by total energy
minimization keeping the lattice parameters fixed at the experimental values.
This was done in a scalar relativistic approximation.
The electronic structures and related properties were then calculated
including spin-orbit. The results are based on dense Brillouin zone
samplings of $\sim$5900 points in the irreducible wedge.
The conductivity anisotropy was determined using the BoltzTraP
code as applied to the electronic structure
including spin-orbit. \cite{boltztrap}

\section*{Acknowledgments}

This work was supported by the Department of Energy, BES, Materials Sciences and Engineering Division.


\bibliography{Ta2PdSe5}

\section*{Figure Legends}

\section*{Tables}
\begin{table}[!ht]
\caption{Structure of Ta$_2$PdSe$_5$ based on relaxation of the
atomic coordinates using the experimental lattice parameters,
spacegroup 12, $C2/m$, $a$=12.801 \AA, $b$=18.7554 \AA, 
$c$= 3.117 \AA, $\gamma$=114.7266$^\circ$. ``b.v." denotes
the bond-valence sum.}
\begin{tabular}{|l|cccc|}
\hline
atom & $x$ & $y$ & $z$ & b.v. \\
\hline
Ta1 & 0.3648 & 0.3348 & 1/2 & 4.71 \\
Ta2 & 0.2307 & 0.1618 & 0   & 5.04 \\
Pd1 & 0.5000 & 0.5000 & 0   & 2.11 \\
Pd2 & 0.0000 & 0.0000 & 0   & 1.84 \\
Se1 & 0.3312 & 0.9618 & 0   & 2.90 \\
Se2 & 0.4043 & 0.2070 & 1/2 & 3.18 \\
Se3 & 0.5412 & 0.3827 & 0   & 2.87 \\
Se4 & 0.7914 & 0.4145 & 1/2 & 2.73 \\
Se5 & 0.6458 & 0.2319 & 0   & 3.09 \\
\hline
\end{tabular}
\label{tab:struct}
\end{table}

\end{document}